\documentclass[conference]{IEEEtran}
\IEEEoverridecommandlockouts
\usepackage{cite}
\usepackage{amsmath,amssymb,amsfonts}
\usepackage{algorithmic}
\usepackage{subcaption}

\usepackage{float}

\usepackage{graphicx}
\usepackage{textcomp}
\usepackage{xcolor}
\usepackage{pgfplots}
\pgfplotsset{compat=1.18} 
\usepackage{filecontents}
\usepackage{amsmath}
\usepackage{amssymb}
\usepackage{algorithm}
\usepackage{algorithmic}
\usepackage{graphicx}
\usepackage{url}

\usepackage{tikz}
\usepackage{lipsum}
\usetikzlibrary{positioning, arrows.meta, calc}
\usepackage{tikz}

\usepackage{xcolor}

\usepackage[bookmarks=false]{hyperref}

\def\BibTeX{{\rm B\kern-.05em{\sc i\kern-.025em b}\kern-.08em
    T\kern-.1667em\lower.7ex\hbox{E}\kern-.125emX}}
    
\begin{document}

\title{Semantic-Aware Sub-Band Allocation for \\ Terahertz Communications\\

\thanks{This work is supported by the American University of Beirut University Research Board and Vertically Integrated Projects Program.}
}

\author{
        Fatima Ismail, Hadi Sarieddeen, and Jihad Fahs
 \\
        Department of Electrical and Computer Engineering\\ American University of Beirut, Beirut 1107 2020, Lebanon\\
        fmi15@mail.aub.edu, hadi.sarieddeen@aub.edu.lb, jihad.fahs@aub.edu.lb
}

\maketitle

\begingroup
\renewcommand\thefootnote{}
\footnotetext{\footnotesize
\copyright~2026 IEEE. Personal use of this material is permitted.
Permission from IEEE must be obtained for all other uses, in any current or
future media, including reprinting/republishing this material for advertising
or promotional purposes, creating new collective works, for resale or
redistribution to servers or lists, or reuse of any copyrighted component
of this work in other works.}
\addtocounter{footnote}{-1}
\endgroup

\begin{abstract}

This paper studies semantic-aware sub-band allocation for terahertz (THz) communication systems, where frequency-selective molecular absorption creates highly non-uniform sub-band qualities. Unlike conventional formulations, semantic fidelity depends nonlinearly on the signal-to-noise ratio (SNR) and is also sentence-specific, leading to a non-separable assignment problem that is generally not solvable using simple ordering-based policies. To address this, we use a sentence-BERT (SBERT)-based surrogate model that predicts semantic fidelity from the sentence embedding and sub-band SNR. We propose an importance-aware scheduler that assigns sentences to sub-bands based on their semantic contribution using an oracle utility function that captures importance-weighted semantic similarity across sentence-sub-band pairs. A neural scheduler is then trained through imitation learning to approximate 
the oracle policy at more than $200\times$ lower runtime than full DeepSC-based oracle evaluation. Integrated with a deep-learning-enabled semantic communication (DeepSC) system, the 
proposed method consistently outperforms all benchmark schemes and approaches 
oracle-level performance under realistic THz channel conditions.

\end{abstract}


\section{Introduction}

Future generations of wireless communication target massive connectivity, ultra-low latency, and extremely high data rates~\cite{9514889}, driven in part by semantic communication, which focuses on transmitting meaning rather than raw bits~\cite{yang2022semantic}, and terahertz (THz) communication~\cite{9112745}, which leverages the 0.1--10~THz band for ultra-high-capacity links (as algorithmic and spectral enablers, respectively). Semantic communication shifts the objective from bit-level accuracy to preserving task-relevant meaning~\cite{yang2022semantic}, enabled by deep learning architectures such as deep learning-enabled semantic communication (DeepSC)~\cite{xie2021deep} and its extensions~\cite{zhang2022unified, weng2021semantic, do2025trangdeepsc, peng2024robust, alhaj2026signdeepsc}. Evaluation increasingly relies on embedding-based metrics such as sentence-BERT (SBERT, bidirectional encoder representations from transformers)~\cite{sbert}, which better capture semantic fidelity than traditional metrics compared to the bilingual evaluation understudy (BLEU) score~\cite{BLEU}.

Recent works have explored resource allocation in semantic communication systems, where transmission is guided by task relevance. For instance,~\cite{9763856} optimizes semantic similarity under communication constraints,~\cite{10538233} proposes importance-aware allocation, and~\cite{zhang2023toward} develops task-oriented scheduling, while age of information (AoI)-based frameworks~\cite{yates2021age} prioritize timely updates. However, these approaches typically rely on simplified channel models (e.g., additive white Gaussian noise (AWGN) or flat fading) or generic multicarrier settings. 

THz communication exhibits strong frequency-dependent attenuation due to molecular absorption~\cite{9112745, han2022molecular}, leading to structured spectral windows with highly heterogeneous sub-band qualities, accurately characterized using the high-resolution transmission molecular absorption (HITRAN) database~\cite{rothman2021history,9591285}. Although sub-band allocation in THz systems has been studied~\cite{shafie2022spectrum}, existing approaches remain bit-centric, optimizing throughput based on physical-layer quantities such as channel conditions and power while ignoring the transmitted content. In contrast, semantic-aware allocation methods that account for content importance typically overlooked the pronounced frequency selectivity of THz channels~\cite{9763856,10538233,zhang2023toward,yates2021age}. This creates a fundamental gap: THz allocation methods ignore semantics, while semantic-aware methods ignore THz spectral heterogeneity.

To the best of our knowledge, semantic-aware resource allocation over frequency-selective THz bands has not yet been explored. We address this gap by proposing a formulation that jointly captures semantic importance and frequency-selective channel quality. This coupling fundamentally alters the problem structure: the utility of assigning a sub-band depends jointly on the sentence and the experienced signal-to-noise ratio (SNR), yielding a non-separable formulation that departs from conventional designs and precludes classical sorting-based solutions. Our main contributions are
\begin{itemize}
    \item We formulate a semantic-aware sub-band allocation problem for THz communication using DeepSC, jointly capturing sentence importance and frequency-selective channel conditions.
    
    \item We show that semantic fidelity is nonlinear in SNR and sentence-dependent, leading to a non-separable assignment problem that cannot, in general, be solved via sorting-based policies. We develop an SBERT-based surrogate model that efficiently approximates semantic fidelity as a function of sentence embeddings and SNR.
    
    \item We design a neural scheduler trained via imitation learning to approximate optimal oracle assignment at low complexity, achieving consistent gains over benchmark schemes under realistic THz channels.
\end{itemize}

\label{sec:sysmodel}

\section{System and Channel models}

We consider an end-to-end semantic communication system based on DeepSC~\cite{xie2021deep}, adopting the original DeepSC architecture, that transmits the semantic content of a sentence $\mathbf{v}=[v_1,\dots,v_L]$. The transmitter comprises a semantic encoder $S_{\beta}(\cdot)$ and a channel encoder $C_{\alpha}(\cdot)$, parameterized by $\beta$ and $\alpha$. The semantic encoder extracts task-relevant features, which the channel encoder maps to a symbol vector $\mathbf{x}=C_{\alpha}(S_{\beta}(\mathbf{v})) \in \mathbb{C}^{M\times1}$ for transmission over a THz channel within a coherence interval of length $M$.

We consider a frequency-selective THz transmission window $[f_{\min}, f_{\max}]$ partitioned into $S$ sub-bands with distinct channel conditions due to molecular absorption. This selectivity is governed by the absorption coefficient $K(f)$, which varies with frequency due to atmospheric gas resonances, causing stronger attenuation near absorption peaks and higher effective SNR within spectral windows. Transmission over sub-band $s$ is modeled as $\mathbf{y}_s = h_s \mathbf{x} + \mathbf{n}, \; s=1,\dots,S$, where $h_s$ is the effective channel gain (including propagation and beamforming), and $\mathbf{n}\!\sim\!\mathcal{CN}(\mathbf{0},\sigma^2\mathbf{I}_M)$ is circularly symmetric complex Gaussian noise vector. The per-symbol SNR is $\gamma_s = \frac{P|h_s|^2}{\sigma^2}$, where $P=\mathbb{E}[|x_m|^2]$ is the average transmit power. At the receiver, the channel and semantic decoders reconstruct the sentence as \(\hat{\mathbf{v}}=S^{-1}_{\chi}(C^{-1}_{\delta}(\mathbf{y}))\).

We generate realistic THz channels using the TeraMIMO simulator~\cite{9591285}, based on a Saleh--Valenzuela clustered multipath model with frequency-dependent propagation and molecular absorption (HITRAN data) under a subarray (SA) architecture. The transmitter and receiver comprise SAs with multiple antenna elements. At subcarrier $k$, the beamformed channel between transmit and receive SAs is~\cite{9591285}
\begin{equation*}
\begin{aligned}
{h}_{q^{(r)},q^{(t)}}[k]
&=
{\mathbf a}^{(r)T}\!\Big(
\boldsymbol{\Phi}_{0}^{(r)}{}_{q^{(r)},\,q^{(r)}}
\Big)\,
\mathbf{H}_{q^{(r)},q^{(t)}}[k] \\
&\quad\times
{\mathbf a}^{(t)}\!\Big(
\boldsymbol{\Phi}_{0}^{(t)}{}_{q^{(t)},\,q^{(t)}}
\Big),
\end{aligned}
\end{equation*}
where $q^{(t)}$ and $q^{(r)}$ index the transmit and receive SAs, respectively,
$\boldsymbol{\Phi}_{0}^{(t)}{}_{q^{(t)},\,q^{(t)}}$ and
$\boldsymbol{\Phi}_{0}^{(r)}{}_{q^{(r)},\,q^{(r)}}$ denote their corresponding
beam-steering directions,
${\mathbf a}^{(t)}(\cdot)$ and ${\mathbf a}^{(r)}(\cdot)$ are the corresponding beamforming vectors,
and $\mathbf{H}_{q^{(r)},q^{(t)}}[k]$ is the frequency-domain multiple-input multiple-output (MIMO) channel between the SAs, capturing THz-specific line-of-sight (LoS) and multipath losses, antenna gains, array responses, and molecular absorption.

\section{Problem Formulation}
\label{sec:problem}

We consider a multi-user semantic communication system where a transmitter serves $K$ users, each providing a sentence $\mathbf{v}$ per slot. The THz window is partitioned into $S$ orthogonal sub-bands with distinct SNRs due to path loss and frequency-selective molecular absorption. Following~\cite{shafie2022spectrum}, we neglect inter-user interference and set $S\!=\!K$ for one-to-one user--sub-band assignment. Each sentence $\mathbf{v}$ is encoded using DeepSC~\cite{xie2021deep}, transmitted over a sub-band with SNR $\gamma_s$, and reconstructed as $\hat{\mathbf{v}}$. Semantic fidelity is measured via SBERT cosine similarity~\cite{sbert}, $m = \cos\big( \mathrm{SBERT}(\mathbf{v}), \mathrm{SBERT}(\hat{\mathbf{v}}) \big)$. Since $\hat{\mathbf{v}}$ depends on both $\gamma_s$ and $\mathbf{v}$, we write $m \triangleq g(\gamma_s,\mathbf{v})$ for some nonlinear function $g$.

\subsection{Semantic Importance and Assignment Model}

In each time slot, we assign each sentence $\mathbf{v}_i$, $1 \leq i \leq K$ an importance score $\pi_i \in [0,1]$ capturing its semantic relevance. We compute $\pi_i$ using token-level importance weights $\{w_{i,\ell}\}_{\ell=1}^{L}$ generated by meaning-aware response scoring (MARS)~\cite{bakman2024mars}. The sentence-level score is obtained via a combination of average and maximum importance,
\begin{equation*}
\pi_i
=
\alpha \frac{1}{L} \sum_{\ell=1}^{L} w_{i,\ell}
+
(1-\alpha)\max_{\ell} w_{i,\ell},
\end{equation*}
where $\alpha \!\in\! [0,1]$ balances global and peak contributions; we set $\alpha\!=\!0.5$, though other values apply depending on the task.

Let $\mathbf{X}\!=\![x_{i,s}]$, $1 \!\leq\! i,s \!\leq\! K$, be a binary assignment matrix, where $x_{i,s}\!=\!1$ if sentence $i$ is assigned to sub-band $s$, and $0$ otherwise. Assuming a one-to-one assignment per time-slot, each sentence is assigned to exactly one sub-band and each sub-band serves exactly one sentence. The semantic utility is
\begin{equation}
U(\mathbf{X})=\sum_{i=1}^{K}\sum_{s=1}^{K}\pi_i\,g(\gamma_s,\mathbf{v}_i)\,x_{i,s},
\label{eq:utility}
\end{equation}
which captures the importance-weighted semantic fidelity achieved under assignment $\mathbf{X}$. The optimal allocation is then,
\begin{align*}
\max_{\mathbf{X}} \quad & U(\mathbf{X}) \\
\text{s.t.} \quad 
& \sum_{s=1}^{S} x_{i,s}=1,\ \forall i,\quad
  \sum_{i=1}^{K} x_{i,s}=1,\ \forall s, \nonumber\\
& x_{i,s}\in\{0,1\}. \nonumber
\end{align*}

\subsection{Structure of the Optimal Assignment}

We next characterize the structure of the optimal solution. 
\begin{figure*}[t]

    \centering
    \includegraphics[width=0.85\linewidth]{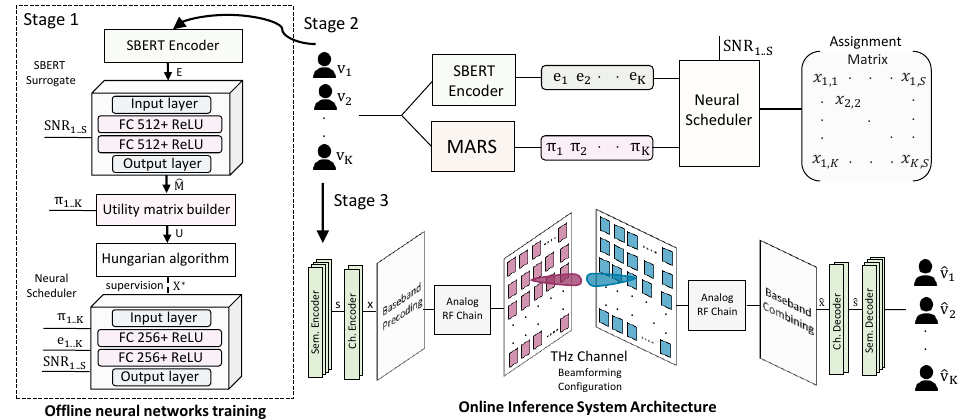}
    \caption{Proposed pipeline including surrogate training, scheduler inference, and DeepSC-based transmission.}
    \label{fig:my_figure}
\end{figure*}
\subsubsection{Pairwise Exchange Analysis}

For a fixed sentence $\mathbf{v}_i$, the semantic similarity $g(\gamma,\mathbf{v}_i)$ is non-decreasing in SNR, yet nonlinear and sentence-dependent. 

We highlight the impact of this relationship using a pairwise comparison example. Consider two sentences $\mathbf{v}_i,\mathbf{v}_j$ and two sub-bands $a,b$ with $\gamma_a > \gamma_b$. Assigning $\mathbf{v}_i\!\to\!a$, $\mathbf{v}_j\!\to\!b$, the utility is
\begin{equation*}
U_{ij} = \pi_i g(\gamma_a,\mathbf{v}_i) + \pi_j g(\gamma_b,\mathbf{v}_j).
\end{equation*}
By swapping the assignment, the utility becomes  
\begin{equation*}
U_{ij}^{\text{swap}} = \pi_i g(\gamma_b,\mathbf{v}_i) + \pi_j g(\gamma_a,\mathbf{v}_j).
\end{equation*}

The current assignment is optimal if $U_{ij} \ge U_{ij}^{\text{swap}}$, yielding
\begin{align*}
\pi_i g(\gamma_a,\mathbf{v}_i) + \pi_j g(\gamma_b,\mathbf{v}_j)
&\ge \pi_i g(\gamma_b,\mathbf{v}_i) + \pi_j g(\gamma_a,\mathbf{v}_j) \\
\Rightarrow\;
\pi_i\big(g(\gamma_a,\mathbf{v}_i) - g(\gamma_b,\mathbf{v}_i)\big)
&\ge \pi_j\big(g(\gamma_a,\mathbf{v}_j) - g(\gamma_b,\mathbf{v}_j)\big).
\label{eq:exchange}
\end{align*}

This indicates that different sentences benefit differently from SNR improvements, so the marginal gain of assigning a higher-quality sub-band is sentence-dependent. As a result, the optimal assignment is driven by importance-weighted marginal gains rather than simple ordering.

\subsubsection{Condition for Greedy Optimality}
We identify a sufficient condition under which greedy sorting is globally optimal. Suppose all sentences share the same semantic response, i.e., $g(\gamma,\mathbf{v}_i)=h(\gamma)$ for all $i$, where $h(\gamma)$ is non-decreasing. 

Let the importance scores and sub-band SNRs be sorted as
\begin{equation*}
\pi_1 \ge \pi_2 \ge \cdots \ge \pi_K,
\qquad
\gamma_1 \ge \gamma_2 \ge \cdots \ge \gamma_K.
\end{equation*}

We claim that the sorted assignment $s = i$ is optimal. Consider any non-sorted assignment. Then there exists $i<j$ such that $\gamma_{i} < \gamma_{j}$ despite $\pi_i \ge \pi_j$. It is clear that swapping the two sub-bands does not decrease utility. Repeated exchanges yield the sorted assignment as globally optimal.

\section{System Architecture}

Fig.~\ref{fig:my_figure} shows the full pipeline: offline surrogate and scheduler training (Stage~1), online assignment inference (Stage~2), and DeepSC-based semantic transmission (Stage~3). In Stage~1, the SBERT-based surrogate predicts semantic similarity from sentence embeddings and SNR, enabling utility-matrix construction from $\{e_i=\text{SBERT}(\mathbf{v}_i)\}$, $\{\pi_i\}$, and $\{\gamma_s\}$. The Hungarian algorithm then generates oracle assignments used to train the scheduler by imitation learning. In Stage~2, the surrogate is discarded, and the scheduler maps embeddings, MARS-based importance scores, and sub-band SNRs to an assignment matrix. In Stage~3, the selected allocation is applied to the DeepSC THz-band transceiver.

\subsection{Model Architectures and Training}

We employ two lightweight neural modules: an SBERT surrogate for semantic fidelity estimation and a neural scheduler for sub-band allocation.

\paragraph{SBERT Surrogate}
The surrogate models the mapping $(\text{sentence},\gamma)\mapsto$ semantic similarity, enabling efficient evaluation of $g(\gamma,\mathbf{v})$ without requiring full DeepSC inference. This approximation is justified by the smooth SNR dependence of semantic similarity, enabling accurate regression from embeddings and channel conditions. The input consists of a 769-dimensional vector (a 768-SBERT embedding concatenated with the SNR), which is processed by a $769 \!\rightarrow\! 512 \!\rightarrow\! 512 \!\rightarrow\! 1$ multi-layer perceptron (MLP) with rectified linear unit (ReLU) activations and dropout of 0.1. The model is trained on Europarl sentences spanning multiple sub-bands and power levels (25-45 dBm) using the Adam optimizer (learning rate $10^{-3}$, batch size $2048$) for 60 epochs, with a mean squared error (MSE) loss.

\paragraph{Neural Scheduler}
The scheduler learns a permutation-valued mapping from sentence semantics, importance, and channel state to sub-band assignments. Its input is $[e_1,\dots,e_K,\pi_1,\dots,\pi_K,\mathrm{SNR}_1,\dots,\mathrm{SNR}_K]$, and it uses an input $\!\rightarrow\! 256 \!\rightarrow\! 256 \!\rightarrow\! K^2\!$ MLP. The output is reshaped into a $K\!\times\! K$ score matrix and converted into a one-to-one assignment by selecting the highest-scoring available sub-band for each sentence
 (conflicts resolved greedily for low-complexity inference). Training uses imitation learning with Hungarian-based targets from the surrogate utility, optimized with Adam (lr $10^{-3}$, batch size $256$) for 100 epochs using row-wise cross-entropy and column regularization ($\lambda_{\mathrm{col}}\!=\!0.1$).

\subsection{Surrogate Error}
The scheduler imitates oracle assignments derived from the surrogate,
whose prediction error depends on training data, test sentences, and
channel realizations. For sentence $i$ and sub-band $s$, let
$e_{i,s} \triangleq \hat{g}_i(\gamma_s)-g_i(\gamma_s)$ denote the surrogate
prediction error. Then the true and surrogate utility terms in the utility sum satisfy
$U_{i,s}=\pi_i g_i(\gamma_s)$ and $\hat{U}_{i,s}=\pi_i\hat{g}_i(\gamma_s)$,
giving the entry-wise perturbation $\hat{U}_{i,s}-U_{i,s}=\pi_i e_{i,s}$, leading the utility error to be directly controlled by the
surrogate prediction error. For any feasible $\mathbf{X}$, the aggregate
perturbation is
\begin{equation*}
\Delta_U(\mathbf{X})
\triangleq
\hat{U}(\mathbf{X})-U(\mathbf{X})
=
\sum_{i=1}^{K}\sum_{s=1}^{K}\pi_i e_{i,s}x_{i,s}.
\label{eq:utility_perturbation}
\end{equation*}
Taking the expected absolute value gives
\begin{align}
\mathbb{E}\!\left[|\Delta_U(\mathbf{X})|\right]
&=
\mathbb{E}\!\left[
\left|
\sum_{i=1}^{K}\sum_{s=1}^{K}\pi_i e_{i,s}x_{i,s}
\right|
\right] \notag\\
&\leq
\sum_{i=1}^{K}\sum_{s=1}^{K}
\pi_i x_{i,s}\mathbb{E}\!\left[|e_{i,s}|\right],
\label{eq:expected_abs_gap_general}
\end{align}
by the triangle inequality and linearity of expectation.

For each sentence and subcarrier pair, define the expected absolute error and its worst-case value as
$\eta_{i,s} \triangleq \mathbb{E}[|e_{i,s}|]$ and
$\eta_{\max} \triangleq \max_{i,s}\eta_{i,s}$, respectively. This avoids
assuming identical error distributions across sentence--sub-band pairs. Using $\sum_s x_{i,s}=1$ for every $i$, \eqref{eq:expected_abs_gap_general}
becomes
\begin{equation}
\mathbb{E}\!\left[|\hat{U}(\mathbf{X})-U(\mathbf{X})|\right]
\leq \eta_{\max}\sum_{i=1}^{K}\pi_i .
\label{eq:expected_utility_gap}
\end{equation}

Let $\mathbf{X}^{\star}$ maximize the true utility $U(\mathbf{X})$, and let
$\hat{\mathbf{X}}^{\star}$ maximize the surrogate utility $\hat{U}(\mathbf{X})$. The
surrogate-induced oracle loss can be decomposed as
\begin{align*}
U(\mathbf{X}^{\star})-U(\hat{\mathbf{X}}^{\star})
&=\!
\big(U(\mathbf{X}^{\star})-\hat{U}(\mathbf{X}^{\star})\big) \!+\!
\big(\hat{U}(\mathbf{X}^{\star})-\hat{U}(\hat{\mathbf{X}}^{\star})\big) \notag\\
&\quad+
\big(\hat{U}(\hat{\mathbf{X}}^{\star})-U(\hat{\mathbf{X}}^{\star})\big).
\end{align*}
The middle term is non-positive since $\hat{\mathbf{X}}^{\star}$ maximizes
$\hat{U}$, and $U(\mathbf{X}^{\star})-U(\hat{\mathbf{X}}^{\star})\geq0$ since
$\mathbf{X}^{\star}$ maximizes $U$. Hence,
\begin{align*}
U(\mathbf{X}^{\star})-U(\hat{\mathbf{X}}^{\star})
&\leq
\left|U(\mathbf{X}^{\star})-\hat{U}(\mathbf{X}^{\star})\right| \notag\\
&\qquad \qquad +
\left|\hat{U}(\hat{\mathbf{X}}^{\star})-U(\hat{\mathbf{X}}^{\star})\right|.
\label{eq:surrogate_loss_abs}
\end{align*}
Taking expectations and applying \eqref{eq:expected_utility_gap} to both
assignments yields
\begin{equation}
\mathbb{E}\!\left[
U(\mathbf{X}^{\star})-U(\hat{\mathbf{X}}^{\star})
\right]
\leq
2\eta_{\max}\sum_{i=1}^{K}\pi_i .
\label{eq:expected_oracle_gap}
\end{equation}
Thus, the expected utility loss resulting from the surrogate-assisted oracle construction is
controlled by the surrogate prediction error, hence linking the surrogate accuracy to
the reliability of the oracle labels used for scheduler training.

\section{Studied Algorithm}

The offline pipeline has two stages: surrogate training (Alg.~\ref{alg:surrogate}) and scheduler training by oracle imitation (Alg.~\ref{alg:training}). The surrogate predicts the semantic similarity $\hat{m}_{i,s}$ for each embedding--SNR pair, avoiding repeated DeepSC+SBERT evaluations. For each frame, it builds the utility matrix $\hat{U}_{i,s}=\pi_i\hat{m}_{i,s}$ from $\{e_i\}$, $\{\pi_i\}$, and $\{\gamma_s\}$, and the oracle assignment is obtained as
$\mathbf{X}^{\mathrm{oracle}}=H(\hat{\mathbf{U}})$,
where $H(\cdot)$ denotes the Hungarian algorithm. The scheduler $f_\theta$ then learns to map $(\mathbf{E},\boldsymbol{\pi},\boldsymbol{\gamma})$ to $\mathbf{X}^{\mathrm{oracle}}$ using cross-entropy, where $\theta$ denotes its trainable parameters. At inference, both the surrogate and Hungarian solver are discarded, and $f_\theta$ outputs assignments in a single forward pass.

\label{sec:algorithm}
\begin{algorithm}[t]

\caption{SBERT Surrogate Training}
\label{alg:surrogate}
\begin{algorithmic}[1]
\STATE \textbf{Input:} $\mathcal{D}=\{(e^{(n)},\gamma^{(n)},m^{(n)})\}_{n=1}^N$, surrogate $\mathcal{M}_{\mathrm{sur}}(\cdot;\phi)$, learning rate $\rho$, epochs $E$, batch size $B$
\FOR{$epoch=1,\ldots,E$}
    \FOR{mini-batch $\mathcal{B}\subset\mathcal{D}$}
        \STATE Form $z^{(n)}=[e^{(n)},\gamma^{(n)}]$ and predict $\hat{m}^{(n)}=\mathcal{M}_{\mathrm{sur}}(z^{(n)};\phi)$
        \STATE $\mathcal{L}=\frac{1}{|\mathcal{B}|}\sum_{n\in\mathcal{B}}(\hat{m}^{(n)}-m^{(n)})^2$, \quad
        $\phi\leftarrow\phi-\rho\nabla_\phi\mathcal{L}$
    \ENDFOR
\ENDFOR
\STATE \textbf{Output:} trained surrogate $\mathcal{M}_{\mathrm{sur}}(\cdot;\phi)$
\end{algorithmic}
\end{algorithm}

\begin{algorithm}[t]
\caption{Scheduler Training}
\label{alg:training}
\begin{algorithmic}[1]
\STATE \textbf{Inputs:} $\{s_i\}$, $\{e_i\}$, $\{\pi_i\}$, $\boldsymbol{\gamma}$, $\mathcal{M}_{\mathrm{sur}}$, $\mathcal{H}$, $f_\theta$
\FOR{$t=1,\ldots,T$}
    \STATE Obtain $\{e_i^{(t)}\}$, $\boldsymbol{\pi}^{(t)}$, $\boldsymbol{\gamma}^{(t)}$
    \STATE \textbf{Step 1:} Surrogate utility:
    $\hat{m}_{i,s}=\mathcal{M}_{\mathrm{sur}}(e_i,\mathrm{SNR}_s)$,
    $\hat{U}_{i,s}=\pi_i\hat{m}_{i,s}$, $\forall i,s$
    \STATE \textbf{Step 2:} Oracle:
    $\mathbf{X}^{\mathrm{oracle}}=\mathcal{H}(\hat{\mathbf{U}})$
    \STATE \textbf{Step 3:} Prediction:
    $\widehat{\mathbf{X}}=f_\theta(\mathbf{E}^{(t)},\boldsymbol{\pi}^{(t)},\boldsymbol{\gamma}^{(t)})$
    \STATE \textbf{Step 4:} Update:
    $\mathcal{L}=\mathrm{CE}(\widehat{\mathbf{X}},\mathbf{X}^{\mathrm{oracle}})$,
    $\theta \leftarrow \theta-\eta\nabla_\theta\mathcal{L}$
\ENDFOR
\STATE \textbf{Output:} trained scheduler $f_\theta$
\end{algorithmic}
\end{algorithm}

\section{Results and Discussion}
We evaluate the proposed importance-aware scheduler using importance-weighted semantic utility and SBERT for the top-10\% most important and bottom-10\% least important sentences. We compare against: (i) random assignment (lower bound), (ii) PiOnly, which ranks sentences by $\pi_i$ and assigns them to a fixed random sub-band order, ignoring THz sub-band heterogeneity, (iii) SNROnly, which ranks sub-bands by $\gamma_s$ and assigns them to the randomized sentence order, ignoring semantic importance, (iv) greedy rank-based matching, which pairs higher-$\pi_i$ sentences with higher-$\gamma_s$ sub-bands, (v) proxy Hungarian, which maximizes
$U_{\mathrm{proxy}}(\mathbf{X})=\sum_{i=1}^{K}\sum_{s=1}^{K}\pi_i\gamma_s x_{i,s}$,
obtained by replacing $g(\gamma_s,\mathbf{v}_i)$ in~\eqref{eq:utility} with $\gamma_s$, where both greedy and proxy Hungarian test whether simple combinations of importance and SNR are sufficient, and (vi) oracle Hungarian (upper bound), which maximizes the true SBERT-based utility
$U(\mathbf{X})=\sum_{i=1}^{K}\sum_{s=1}^{K}\pi_i g(\gamma_s,\mathbf{v}_i)x_{i,s}$.

\textit{Simulation settings:} We consider an indoor THz scenario over $[f_{\min}, f_{\max}]$ including molecular absorption peaks (380--450 GHz), with link distance $d\!=\!5$ m. The band is partitioned into $S = K$ sub-bands, each experiencing frequency-dependent attenuation due to molecular absorption, with path-loss exponent equal to $2$~\cite{9473756}. Here, $K$ reflects the number of simultaneously scheduled users per frame; the scheduler is re-applied each frame, making it suitable for arbitrarily large user populations over time. Owing to the short THz wavelength, antenna elements are densely packed with half-wavelength spacing~\cite{8311993}, enabling ultra-massive $4096 \!\times\! 4096$ arrays at both transmitter and receiver~\cite{faisal2020ultramassive} and significant beamforming gains to mitigate path loss and absorption. The channel follows a clustered multipath model with Poisson-distributed clusters and Gaussian angular spreads. Simulations use Europarl (80k/10k train/test sentences).

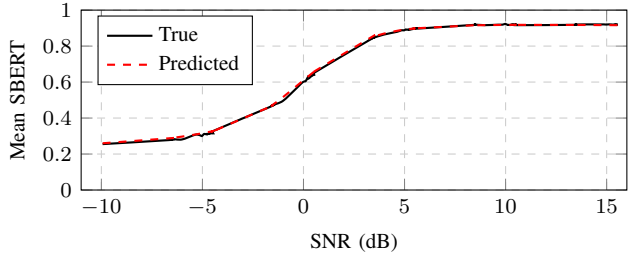
\begin{figure}[t]
\centering
\begin{tikzpicture}
\begin{axis}[
    width=\columnwidth,
    height=0.45\columnwidth,
    xlabel={SNR (dB)},
    ylabel={Mean SBERT},
    xmin=-11, xmax=16,
    ymin=0, ymax=1.0,
    grid=both,
    grid style={dashed,gray!40},
    legend style={
        at={(0.03,0.97)},
        anchor=north west,
        font=\footnotesize,
        draw=black,
        fill=white
    },
    tick label style={font=\footnotesize},
    label style={font=\footnotesize},
    legend cell align={left},
    line width=0.4pt,
]

\addplot[
    smooth,
    thick,
    color=black
]
table[
    col sep=comma,
    x=snr_db,
    y=mean_true_sbert
]{per_snr_metrics.csv};
\addlegendentry{True}

\addplot[
    smooth,
    thick,
    dashed,
    color=red
]
table[
    col sep=comma,
    x=snr_db,
    y=mean_pred_sbert
]{per_snr_metrics.csv};
\addlegendentry{Predicted}

\end{axis}
\end{tikzpicture}
\caption{Average true $\mathbb{E}\left[g_i(\gamma_s)\right]$ and predicted $\mathbb{E}\left[\hat{g}_i(\gamma_s)\right]$ SBERT versus SNR $\gamma_s$.}
\label{fig:surrogate_snr}
\end{figure}

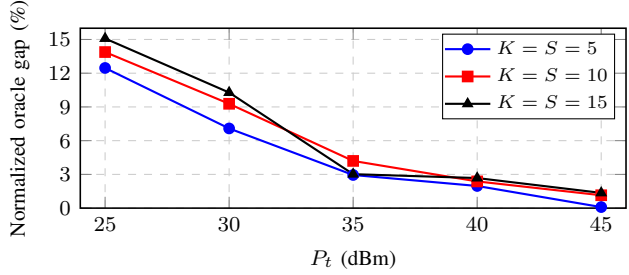
\begin{figure}[t]
\centering
\begin{tikzpicture}
\begin{axis}[
    width=\columnwidth,
    height=0.45\columnwidth,
    xlabel={$P_t$ (dBm)},
    ylabel={Normalized oracle gap (\%)},
    xmin=24, xmax=46,
    ymin=0, ymax=16,
    xtick={25,30,35,40,45},
    ytick={0,3,6,9,12,15},
    grid=both,
    grid style={dashed,gray!40},
    tick label style={font=\footnotesize},
    label style={font=\footnotesize},
    legend style={
        at={(0.98,0.97)},
        anchor=north east,
        font=\scriptsize,
        draw=black,
        fill=white,
        line width=0.4pt,
        cells={anchor=west},
        inner xsep=0.5pt,
        inner ysep=0.5pt
    },
    line width=0.6pt,
    mark size=1.8pt
]

\addplot[
    blue,
    thick,
    mark=*,
]
coordinates {
    (25,12.4659)
    (30,7.0862)
    (35,2.9470)
    (40,1.9705)
    (45,0.0930)
};
\addlegendentry{$K=S=5$}

\addplot[
    red,
    thick,
    mark=square*,
]
coordinates {
    (25,13.8809)
    (30,9.2800)
    (35,4.1978)
    (40,2.3767)
    (45,1.1457)
};
\addlegendentry{$K=S=10$}

\addplot[
    black,
    thick,
    mark=triangle*,
]
coordinates {
    (25,15.0868)
    (30,10.2800)
    (35,3.0198)
    (40,2.6743)
    (45,1.3546)
};
\addlegendentry{$K=S=15$}

\end{axis}
\end{tikzpicture}
\caption{Empirical estimate of the normalized surrogate-induced oracle gap,
$\frac{\mathbb{E}[U(\mathbf{X}^{\star})-U(\hat{\mathbf{X}}^{\star})]}
{\mathbb{E}[U(\mathbf{X}^{\star})]}\times 100\%$,
versus transmit power $P_t$.}
\label{fig:eq16_validation}
\end{figure}

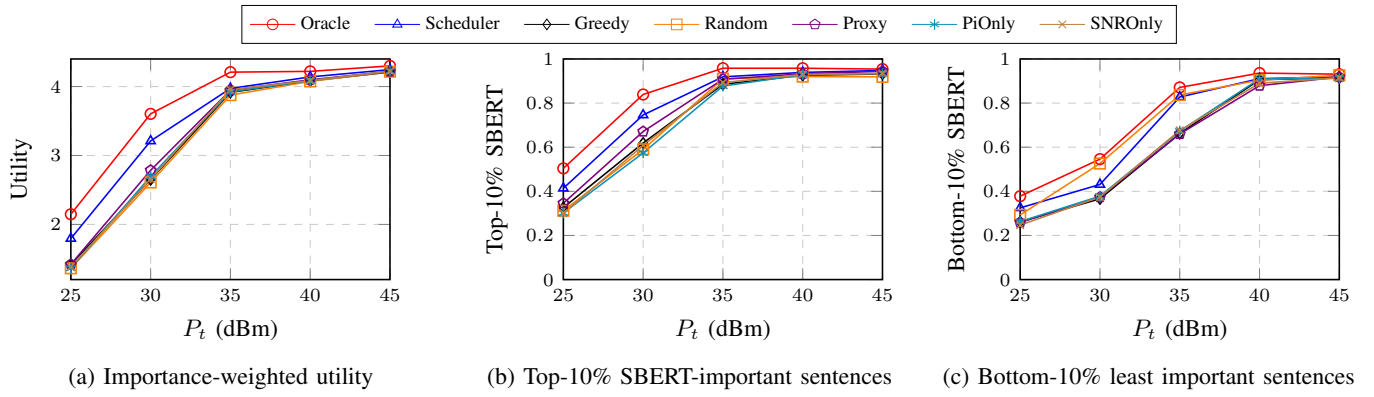
\begin{figure*}[t]
\centering

% -------- (a) Utility --------
\begin{subfigure}[t]{0.32\textwidth}
\centering
\begin{tikzpicture}
\begin{axis}[
    width=\textwidth,
    height=4.5cm,
    xlabel={$P_t$ (dBm)},
    ylabel={Utility},
    grid=both,
    grid style={dashed,gray!40},
    tick label style={font=\scriptsize},
    label style={font=\small},
    line width=0.6pt,
    mark size=2pt,
    every mark/.append style={fill=none, line width=0.2pt}, % <-- force all markers empty
    xmin=25, xmax=45,
    ymin=1.2, ymax=4.4,
    xtick={25,30,35,40,45},
    ytick={0,1,2,3,4,5},
    legend style={
        at={(2,1.07)},
        anchor=south,
        font=\scriptsize,
        draw=black,        % <-- add box
        fill=white,
        line width=0.1pt,
        legend columns=7,
        /tikz/every even column/.append style={column sep=0.25cm}
    }
]

% Oracle
\addplot[red, mark=o] coordinates {
    (25,2.1474)
    (30,3.6038)
    (35,4.2088)
    (40,4.2199)
    (45,4.2996)
};

% Scheduler
\addplot[blue, mark=triangle] coordinates {
    (25,1.7954)
    (30,3.2101)
    (35,3.9724)
    (40,4.1393)
    (45,4.2456)
};

% Greedy
\addplot[black, mark=diamond] coordinates {
    (25,1.4177)
    (30,2.6504)
    (35,3.9151)
    (40,4.0829)
    (45,4.2095)
};

% Random
\addplot[orange, mark=square] coordinates {
    (25,1.3635)
    (30,2.6086)
    (35,3.8758)
    (40,4.0748)
    (45,4.2161)
};

% Proxy
\addplot[violet, mark=pentagon] coordinates {
    (25,1.4210)
    (30,2.7853)
    (35,3.9471)
    (40,4.0978)
    (45,4.2193)
};

% PiOnly
\addplot[cyan!70!black, mark=asterisk] coordinates {
    (25,1.3651)
    (30,2.6973)
    (35,3.9238)
    (40,4.0844)
    (45,4.2230)
};

% SNROnly
\addplot[brown, mark=x] coordinates {
    (25,1.3835)
    (30,2.6564)
    (35,3.9553)
    (40,4.0865)
    (45,4.2243)
};

\legend{Oracle, Scheduler, Greedy, Random, Proxy, PiOnly, SNROnly}

\end{axis}
\end{tikzpicture}
\caption{Importance-weighted utility}
\end{subfigure}\hspace{0.015\textwidth}
% -------- (b) Top-10% --------
\begin{subfigure}[t]{0.32\textwidth}
\centering
\begin{tikzpicture}
\begin{axis}[
    width=\textwidth,
    height=4.5cm,
    xlabel={$P_t$ (dBm)},
    ylabel={Top-10\% SBERT},
    grid=both,
    grid style={dashed,gray!40},
    tick label style={font=\scriptsize},
    every mark/.append style={fill=none, line width=0.2pt},
    label style={font=\small},
    line width=0.6pt,
    mark size=2pt,
    xmin=25, xmax=45,
    ymin=0, ymax=1.0,
    xtick={25,30,35,40,45},
    ytick={0, 0.2,0.4,0.6,0.8,1.0}
]

% Oracle
\addplot[red, mark=o] coordinates {
    (25,0.504)
    (30,0.839)
    (35,0.958)
    (40,0.958)
    (45,0.954)
};

% Scheduler
\addplot[blue, mark=triangle] coordinates {
    (25,0.41394)
    (30,0.746)
    (35,0.919)
    (40,0.939)
    (45,0.948)
};

% Greedy
\addplot[black, mark=diamond] coordinates {
    (25,0.326)
    (30,0.619)
    (35,0.885)
    (40,0.926)
    (45,0.932)
};

% Random
\addplot[orange, mark=square] coordinates {
    (25,0.312)
    (30,0.590)
    (35,0.9100)
    (40,0.920)
    (45,0.919)
};

% Proxy
\addplot[violet, mark=pentagon] coordinates {
    (25,0.346)
    (30,0.671)
    (35,0.908)
    (40,0.933)
    (45,0.944)
};

% PiOnly
\addplot[cyan!70!black, mark=asterisk] coordinates {
    (25,0.302)
    (30,0.574)
    (35,0.877)
    (40,0.929)
    (45,0.934)
};

% SNROnly
\addplot[brown, mark=x] coordinates {
    (25,0.301)
    (30,0.608)
    (35,0.893)
    (40,0.930)
    (45,0.934)
};

\end{axis}
\end{tikzpicture}
\caption{Top-10\% SBERT-important sentences}
\end{subfigure}\hspace{0.015\textwidth}%
% -------- (c) Bottom-10% --------
\begin{subfigure}[t]{0.32\textwidth}
\centering
\begin{tikzpicture}
\begin{axis}[
    width=\textwidth,
    height=4.5cm,
    xlabel={$P_t$ (dBm)},
    ylabel={Bottom-10\% SBERT},
    grid=both,
    grid style={dashed,gray!40},
    tick label style={font=\scriptsize},
    label style={font=\small},
    line width=0.6pt,
    every mark/.append style={fill=none, line width=0.2pt},
    mark size=2pt,
    xmin=25, xmax=45,
    ymin=0, ymax=1.0,
    xtick={25,30,35,40,45},
    ytick={0,0.2,0.4,0.6,0.8,1.0}
]

% Oracle
\addplot[red, mark=o] coordinates {
    (25,0.378)
    (30,0.546)
    (35,0.871)
    (40,0.936)
    (45,0.931)
};

% Scheduler
\addplot[blue, mark=triangle] coordinates {
    (25,0.325)
    (30,0.431)
    (35,0.82746)
    (40,0.9107)
    (45,0.919)
};

% Greedy
\addplot[black, mark=diamond] coordinates {
    (25,0.262)
    (30,0.366)
    (35,0.662)
    (40,0.906)
    (45,0.915)
};

% Random
\addplot[orange, mark=square] coordinates {
    (25,0.293)
    (30,0.526)
    (35,0.837)
    (40,0.904)
    (45,0.926)
};

% Proxy
\addplot[violet, mark=pentagon] coordinates {
    (25,0.259)
    (30,0.374)
    (35,0.660)
    (40,0.879)
    (45,0.918)
};

% PiOnly
\addplot[cyan!70!black, mark=asterisk] coordinates {
    (25,0.263)
    (30,0.378)
    (35,0.673)
    (40,0.910)
    (45,0.912)
};

% SNROnly
\addplot[brown, mark=x] coordinates {
    (25,0.247)
    (30,0.374)
    (35,0.676)
    (40,0.890)
    (45,0.915)
};

\end{axis}
\end{tikzpicture}
\caption{Bottom-10\% least important sentences}
\end{subfigure}
\caption{Semantic-aware sub-band allocation vs. transmit power for $K = S = 15$. }
\label{fig:combined_results}
\end{figure*}

\subsection{Surrogate Accuracy}

Fig.~\ref{fig:surrogate_snr} shows that the surrogate closely tracks the true mean SBERT similarity across SNR. Using a test set of $56{,}250$ samples, it achieves near-zero mean error, $\widehat{\mu}_e=-2.45\times10^{-4}$, with mean absolute error (MAE) $\widehat{\eta}=5.33\times10^{-2}$. Since the utility perturbation satisfies $\hat{U}_{i,s}-U_{i,s}=\pi_i e_{i,s}$ with $0\leq\pi_i\leq1$, this small prediction error implies limited typical distortion in the surrogate utility matrix. Although the empirical worst-case error $\eta_{\max}\approx0.20$ is larger due to outlier sentence--sub-band pairs at very low SNR, the much smaller average error indicates that the bound in~\eqref{eq:expected_oracle_gap} is conservative.

We report in Fig.~\ref{fig:eq16_validation} the empirical normalized oracle gap. For each test frame, $X^\star$ is obtained from the true SBERT utility matrix, while $\hat{X}^\star$ is obtained from the surrogate-predicted matrix; both are evaluated using the true utility. The gap is largest at low transmit power and decreases rapidly with $P_t$, showing that the surrogate-based oracle approaches the true oracle in the reliable semantic operating regime.

\subsection{Importance-Weighted Semantic Utility}

We compare and report the utility of all allocation policies when $K=15$. The results for $K=\{5,10\}$ show exactly similar trends and are omitted for visual clarity and space considerations.

Fig.~\ref{fig:combined_results}(a) shows that  the proposed scheduler outperforms all practical baselines and closely approaches the oracle across all $P_t$, with the largest gains at low $P_t$, where frequency selectivity and nonlinear semantic degradation make allocation critical. It effectively leverages semantic importance and channel heterogeneity, unlike PiOnly, SNROnly, and heuristic methods. As $P_t$ increases, all methods converge due to semantic saturation at high SNR; The scheduler retains a consistent advantage, highlighting the importance of modeling sentence-dependent semantic responses, particularly in low-SNR regimes where naive policies fail.

Fig.~\ref{fig:combined_results}(b) reports semantic fidelity for the top-10\% sentences. The proposed scheduler achieves near-oracle performance in all $P_t$, with the greatest gains at low power, confirming effective prioritization of critical content. Baselines do not consistently align importance with channel quality: SNROnly and greedy rely on channel ordering, PiOnly ignores channel variation, and the proxy is limited by its linear approximation. For the bottom-10\%, as shown in Fig.~\ref{fig:combined_results}(c), the scheduler's performance remains acceptable, indicating a favorable balance between prioritization and fairness. The sub-optimal performance at low $P_t$ can be explained by allocating better resources to important sentences. As $P_t$ increases, all methods converge due to semantic saturation; the scheduler consistently preserves both prioritization and fairness.

\subsection{Scalability Analysis}
Table~\ref{tab:scaling_latency} evaluates scalability for $K\!=\!S\!\in\!\{5,10,15\}$. Averaged over $P_t \!\in\! [25, 45]$ dBm, the scheduler maintains a utility gap below $0.25\%$ relative to the surrogate-assisted Hungarian oracle across all tested $K$; the learned policy generalizes across problem sizes. The scheduler assignment
latency remains stable near $0.49$~ms across all $K$, while the surrogate-assisted
Hungarian latency is $2$--$3\times$ higher; significant at a scale. This gap is expected to widen further at larger $K$, making the
single-pass scheduler increasingly attractive as the number of users and sub-bands
grows. These latency values reflect online assignment only; the full DeepSC oracle requires repeated DeepSC inference for true utility construction and is used only as an upper-bound reference in Fig. \ref{fig:combined_results}.
\begin{table}[t]
\centering
\caption{The proposed scheduler versus the surrogate-assisted Hungarian oracle for different values of $K$.}
\label{tab:scaling_latency}
\small
\setlength{\tabcolsep}{5pt}
\begin{tabular}{c|ccc}
\hline
$K$ & Utility Gap & Sched. Latency & Surr.+Hung. Latency \\
& (\%) & (ms) & (ms) \\
\hline
5  & 0.197 & 0.491 & 1.044 \\
10 & 0.092 & 0.487 & 1.020 \\
15 & 0.239 & 0.490 & 1.319 \\
\hline
\end{tabular}
\end{table}
\subsection{Latency Analysis}
Runtime is measured on a single NVIDIA V100 GPU (32~GB, v100d32q) over the full $10$k-sentence test set. Simple heuristics are faster than the proposed scheduler, with greedy, proxy, SNROnly, PiOnly, and random requiring only $0.025\%$, $0.031\%$, $0.010\%$, $0.0036\%$, and $0.017\%$ of its runtime, respectively. However, oracle construction is over $200\times$ slower because it requires explicit utility-matrix construction and combinatorial assignment. The trained scheduler avoids online Hungarian optimization and produces the assignment through a single feed-forward pass, providing a practical trade-off: higher cost than simple heuristics; orders-of-magnitude lower complexity than the oracle; near-oracle performance.

\section{Conclusion}

We propose a semantic-aware sub-band scheduler for THz communication, integrating DeepSC with a lightweight neural policy trained via surrogate-assisted imitation learning. An SBERT-based surrogate enables efficient fidelity estimation, achieving near-oracle performance at low complexity. Results show that aligning semantic importance with sub-band quality significantly improves reliability in frequency-selective THz channels, underscoring the need for content-dependent resource allocation beyond traditional designs. This highlights a shift from bit-centric to content-aware resource allocation, with future extensions incorporating more physically consistent models and physical-layer reconfigurability~\cite{dkhan2026physically}.

\bibliographystyle{IEEEtran}  
\bibliography{references}   
\end{document}